\documentstyle[11pt,newpasp,twoside]{article}
\def\edcomment#1{\iffalse\marginpar{\raggedright\sl#1\/}\else\relax\fi}
\reversemarginpar

\begin{document}
\title{The Search for Brown Dwarfs around White Dwarfs}
 \author{J. Farihi, E. E. Becklin, B. Zuckerman}
\affil{University of California Los Angeles, 8371 Mathematical Sciences
Building, Los Angeles CA 90095-1562}

\begin{abstract}
The infrared search for substellar companions to nearby white dwarfs has been going for
a little more than a decade.  The most recent phase has been a wide field proper motion
search carried out primarily at Steward Observatory, where we are complete down to $J=18$.
Earlier phases included near field searches at the IRTF and Keck Observatory.  In the last
year we have discovered ten previously unrecognized faint proper motion companions.  Of the
recent discoveries, most are white dwarfs and a few M dwarfs.  GD165B, discovered in 1988
as part of our program, is still the only known companion to a white dwarf with spectral
type later than M.  
\end{abstract}

\section{Description of the Current Survey}
We are conducting a proper motion survey at $1.25\mu$m for low mass stellar and 
substellar companions to nearby white dwarfs.  The advantage of such a search is that
white dwarfs are faint in the infrared, enabling discovery of brown dwarf companions
even relatively close to the primary (Zuckerman \& Becklin 1987 \& 1992; Becklin \&
Zuckerman 1988).  The current wide field survey has been conducted mainly at the Steward
Observatory beginning in 1991.  There on the 2.3m Bok telescope we have used a NICMOS
array and a camera developed by Rieke \& Rieke (Rieke et al. 1993).  With a 3 square
arcminute field of view, we can detect faint companions $2-90''$ from the primary white
dwarf.  We acquire five dithered 90 second images at $J$ band which enables detection
of objects as faint as $J=19$ in the best observing conditions.  

Generally we acquire two epochs for each field surrounding the primary white dwarf.
The typical white dwarf in our sample has a small known proper motion around $0.1-0.2''$/yr.
Since the mid 1990's, the white dwarfs will have moved $\sim0.5-1.0''$ relative to
background stars and galaxies.  We have found that when there are more than 5 objects
which appear in both epoch fields, the scatter in their measured positions is approximately
$0.2''$.  Thus, we can easily detect a $0.5''$  displacement of the primary relative
to background objects in the field.  Hence with a baseline $\geq5$ yrs between epochs,
companions that move with the primary will stand out well against background objects.

The white dwarfs chosen for our survey come from the catalog of McCook \& Sion (1999).
Our targets consist mainly of white dwarfs with modest known proper motions selected
from the Lowell \& Luyten surveys along with some objects from the Palomar-Green survey.
The reason for choosing white dwarfs with smaller proper motions is that stars with
smaller U,V space velocities are statistically more likely to be members of the young
disk (Eggen 1996) and brown dwarfs are brightest when they are young.  White dwarfs are
evolved but not necessarily old; a $3M_{\odot}$ A0 star can evolve into a white dwarf in
only 0.6 Gyr.  Hence by selecting slow movers there is a greater likelihood that we are
looking around younger white dwarfs.  The distance to a typical white dwarf in our sample
of $\sim240$ targets is around 50 pc.

\section{Discoveries}
GD165B was discovered in the very first incarnation of the search for brown dwarfs
around white dwarfs conducted by Becklin \& Zuckerman (1988).  It still remains the only
known possible brown dwarf companion to a white dwarf and the only known companion to a
white dwarf with spectral type later than M.  Zuckerman \& Becklin (1992) discovered or
measured over 20 M dwarf companions to white dwarfs in near field searches conducted at
the IRTF and UKIRT of about 200 white dwarf primaries.

In the current wide field survey, approximately 100 of our white dwarfs have been analyzed
and we have found 10 previously unrecognized proper motion companions to their respective
primaries.  Hence at present we are finding about 10\% of our sample have wide companions.  Three of these companions are likely M dwarfs and the other seven are either confirmed or
likely white dwarfs.  It appears that none of these companions are substellar candidates.
Thus we conclude that not only is there a brown dwarf desert around white dwarfs at close
separations but our data analyzed so far indicates that this trend likely persists at wider
separations as well.

\section{Implications for Star/Brown Dwarf Formation}
One possible explanation for the apparent deficit of substellar companions to white
dwarfs is that binary star formation may favor equally massive components.  There exist
models which favor two accreting protostellar components in a binary system equilibrating
in mass as they form.  One model even predicts that unequal mass binaries are more likely
to form at wide separations due to accretion issues (Bate 2000).  Yet there are many known
WD/RD pairs at close as well as wide separations (Zuckerman \& Becklin 1992;
Silvestri et al. 2001).  These systems would have had very unequal mass components when
on the zero age main sequence.  Radial velocity searches indicate that brown dwarf
companions to main sequence stars are rare at separations of less than 5 AU.  Results
from 2MASS and other studies indicate that brown dwarf companions may be more common
at wide separations (Gizis et al. 2001).  

Companions to white dwarfs suffered a change in their orbital separations since the
time when the primary was on the main sequence.  For companions separated by less than
a few AU when the host star ascends the AGB common envelope interaction will cause the
orbit to decay into an even closer system.  For companions outward of 3 AU, the orbit
will expand by a factor of $M_{\rm ms}/M_{\rm wd}$ (Jeans 1924).  Thus one can expect
wide binary systems with one white dwarf component to have separations that are roughly
2 to 7 times larger than when both components were on the main sequence.  A system with
a white dwarf and a widely separated brown dwarf companion may be vulnerable to disintegration
through close encounters with nearby molecular clouds or other stellar systems over a few
billion years due to the low mass nature of brown dwarfs.

Because brown dwarfs continue to grow dimmer as they age, we must consider the limits of
our sensitivity.  The ages of white dwarfs are still being refined and the refinements
tend to make them older than previous models predicted.  For example, the difference in
predicted cooling times of a typical mass white dwarf with a pure carbon core versus a
pure oxygen core can be as much as 2 Gyr.  What this means for our search is that it
is possible that the ages of white dwarfs could be pushing brown dwarf luminosities past
our sensitivity limits into the late T dwarf regime and beyond.  Based on their space
motions and model predicted cooling times however, the total age of a typical white dwarf
in our sample is $1-3$ Gyr, assuming average masses.  Hence if the models are correct,
we don't expect the age of our sample to play a factor in detecting brown dwarfs down to
about 30 $M_{\rm J}$ for the youngest white dwarfs in our sample.

\section{Conclusions}
The search for brown dwarfs around white dwarfs is of continuing interest for several
reasons.  One is that do not yet know which temperature / spectral class is the lower
cut off for a minimum mass star.  Until we start to discover and study more companions
which straddle and cross the stellar / substellar boundary, the model predicted minimum
temperature value is the only way we can judge whether an object is a brown dwarf or a
low mass star if an accurate age is not known.  Second, by searching for M, L, and T-type
companions to white dwarfs, we are probing ages when brown dwarfs have cooled significantly
and low mass stars of course have not.  When our search or future searches begin to find
low temperature (L-type) companions around old stars such as some white dwarfs, we can then
better constrain stellar / substellar cooling model temperatures, ages and masses.  Also by
conducting this kind of search we want to shed light on the formation of companions to
intermediate mass stars, the progenitors of the white dwarfs we see today.


\begin{references}
\reference Bate, M. R. 2000, \mnras, 314, 33
\reference Becklin, E. E., \& Zuckerman, B. 1988, Nature, 336, 656
\reference Eggen, O. J., 1996, \aj 111, 466  
\reference Gizis, J. E., Kirkpatrick, J. D., Burgasser, A., Reid, I. N., Monet, D. G.,
 		Liebert, J., \& Wilson, J. C. 2001, \apj, 551, L163
\reference Jeans, J. H. 1924. \mnras, 85, 2
\reference McCook, G. P., \& Sion, E. M. 1999, \apjs, 121, 1
\reference Rieke, M. J., Rieke, G. H., Green, E. M., Montgomery, E. F.,
	 	\& Thompson, C. L. 1993, SPIE, 1946, 179
\reference Silvestri, N. M., Oswalt, T. D., Wood, M. A., Smith, J. A., Reid, I. N., Sion,
 		Edward M. 2001, \aj, 121, 503
\reference Zuckerman, B., \& Becklin, E. E. 1987, \apj, 319, L99
\reference Zuckerman, B., \& Becklin, E. E. 1992, \apj, 386, 260
\end{references}
\end{document}